\documentclass[twocolumn,showpacs,prb]{revtex4}

\usepackage{graphicx}
\usepackage{dcolumn}
\usepackage{bm}

\begin{document}

\preprint{APS/123-QED}

\title{Systematic Characterization of Upper Critical Fields for MgB$_2$ Thin Films 
by means of the Two-Band Superconducting Theory 
}

\author{Satoru Noguchi}
\altaffiliation[Also at ]{Institute for Nanofabrication Research, Osaka Prefecture University, 1-1 Gakuen-cho, Naka-ku, Sakai, Osaka 599-8531, Japan.}
\email{noguchi@pe.osakafu-u.ac.jp}
\author{Akihiro Kuribayashi}
\author{Takekazu Ishida}
\altaffiliation[Also at ]{Institute for Nanofabrication Research, Osaka Prefecture University, 1-1 Gakuen-cho, Naka-ku, Sakai, Osaka 599-8531, Japan.}
\affiliation{Department of Physics and Electronics, Osaka Prefecture University, 1-1 
Gakuen-cho, Naka-ku, Sakai, Osaka 599-8531, Japan.}

\author{Tatsunori Oba}
\author{Hiroki Iriuda}
\author{Masato Yoshizawa}
\altaffiliation[Also at ]{JST Satellite Iwate, 3-35-2 Iioka-shinden, Morioka, Iwate 020-0852, Japan.}
\affiliation{Graduate School of Engineering, Iwate University, 4-3-5 Ueda, Morioka, Iwate 020-8551, Japan.}

\author{Yoshitomo Harada}
\affiliation{JST Satellite Iwate, 3-35-2 Iioka-shinden, Morioka, Iwate 020-0852, Japan.}

\author{Shigehito Miki}
\altaffiliation[Also at ]{Institute for Nanofabrication Research, Osaka Prefecture University, 1-1 Gakuen-cho, Naka-ku, Sakai, Osaka 599-8531, Japan.}
\altaffiliation[Also at ]{CREST, Japan Science and Technology Agency, 4-1-8, Honcho, Kawaguchi, Saitama 332-0012, Japan.}
\author{Hisashi Shimakage}
\altaffiliation[Also at ]{Institute for Nanofabrication Research, Osaka Prefecture University, 1-1 Gakuen-cho, Naka-ku, Sakai, Osaka 599-8531, Japan.}
\author{Zhen Wang}
\altaffiliation[Also at ]{Institute for Nanofabrication Research, Osaka Prefecture University, 1-1 Gakuen-cho, Naka-ku, Sakai, Osaka 599-8531, Japan.}
\affiliation{National Institute of Information and Communications Technology, 588-2 Iwaoka-cho, Nishi-ku, Kobe, Hyogo 651-2429, Japan.}

\author{Kazuo Satoh}
\altaffiliation[Also at ]{Institute for Nanofabrication Research, Osaka Prefecture University, 1-1 Gakuen-cho, Naka-ku, Sakai, Osaka 599-8531, Japan.}
\author{Tsutomu Yotsuya}
\altaffiliation[Present Address: ]{Nanoscience and Nanotechnology Research Center, Osaka Prefecture University, 1-1 Gakuen-cho, Naka-ku, Sakai, Osaka 599-8531, Japan.}
\altaffiliation[Also at ]{Institute for Nanofabrication Research, Osaka Prefecture University, 1-1 Gakuen-cho, Naka-ku, Sakai, Osaka 599-8531, Japan.}
\affiliation{Technology Research Institute of Osaka Prefecture, 2-7-1 Ayumino, Izumi, Osaka 594-1157, Japan.}

\date{\today}

\begin{abstract}

We present experimental results of the upper critical fields $H_{\rm c2}$ of various MgB$_2$ thin films prepared by the molecular beam epitaxy, multiple-targets sputtering, and co-evaporation deposition apparatus. 
Experimental data of the $H_{\rm c2}(T)$ are successfully analyzed by applying the Gurevich theory of dirty two-band superconductivity in the case of $D_{\pi}/D_{\sigma}>1$, where $D_{\pi}$ and $D_{\sigma}$ are the intraband electron diffusivities for $\pi$ and $\sigma$ bands, respectively. 
We find that the parameters obtained from the analysis are strongly correlated to the superconducting transition temperature $T_{\rm c}$ of the films. 
We also discuss the anormalous narrowing of the transition width at intermediate temperatures confirmed by the magnetoresistance measurements. 

\end{abstract}


\maketitle

\section{Introduction}
Recently, multiband superconductivity has attracted much attention since a novel metallic superconductor MgB$_2$ discovered at the beginning of the 21st century \cite{nagamatsu2001} has been revealed not only to show the highest $T_{\rm c}$ among intermetallic compounds but also to be a prototype two-band superconductor by numerous experimental and theoretical studies \cite{choi2002nature, choi2002PRB, golubov2002, golubov2003, gurevich2003, gurevich2004, braccini2005, gurevich2007}. 
The superconductivity of MgB$_2$ occurs in both $\sigma$ and $\pi$ bands of the hexagonal B layer which has a similar electronic structure to a graphite \cite{kortus2001, liu2001, choi2002nature, choi2002PRB}. 
The $\sigma$ band, contributing to a chemical bond between B atoms, is two-dimensional orbital localized in the $ab$-plane of the hexagonal structure. 
On the other hand, the $\pi$ band is an antibonding orbital along the $c$-axis, which spread out three-dimensionally. 
Therefore, the interaction between $\sigma$ and $\pi$ bands is very weak and then two superconducting gaps open below the superconducting transition temperature $T_{\rm c}$ in both bands. 

Two-band superconducting character in MgB$_2$ appears in temperature dependence of the upper critical field $H_{\rm c2}$, which shows an upward curvature below $T_{\rm c}$. 
Superconducting anisotropy parameter $\gamma$ determined from the ratio of $H_{\rm c2}$ for the $ab$-direction to that for the $c$-direction shows also temperature dependence. 
The $\gamma$ value for the single crystal increases with decreasing temperature \cite{lyard2002}. 
Drastic enhancement of the $H_{\rm c2}$ is observed in dirty samples such as the nonmagnetic impurity-doped MgB$_2$ bulks and/or films, where $\gamma$ becomes small with showing a variety of the temperature dependence \cite{jung2001, ferrando2003, gurevich2004, noguchi2005, braccini2005}: in a case of a high resistivity film, the $\gamma$ reduces with decreasing temperature \cite{gurevich2004}. 
These behaviors are not explained by only the dimensional crossover as usually discussed in the layered superconductors \cite{lawrence1971}, but the superimpose of the $H_{\rm c2}$ with two distinct superconducting gap \cite{gurevich2007, golubov2003}. 

There are mainly two disorder effects due to nonmagnetic impurity and/or crystal imperfection on the superconducting properties: one is suppression of $T_{\rm c}$ by weakening the superconducting coupling and the other is enhancement of the initial slope of the $H_{\rm c2}$ at ${T_{\rm c}}$ by reducing the electron mean free path \cite{helfand1964}. 
In the two-band superconductor, there is very clever situation in which the great intraband scattering raises the $H_{\rm c2}$ and the small interband scattering affects little suppression on $T_{\rm c}$. 
The shape of the $H_{\rm c2}(T)$ curve is changed depending on which intraband scattering in the $\sigma$ or $\pi$ band is large. 
Gurevich gave the calculation formula for $H_{\rm c2}(T)$ curves in the framework of dirty two-band superconductivity \cite{gurevich2003, gurevich2007}. 
He calculated the $H_{\rm c2}(T)$ curves by using the parameters of the interband scattering $g$ and the diffusivities $D_{\sigma}$ and $D_{\pi}$ in the $\sigma$ and $\pi$ bands including the anisotropy~\cite{gurevich2007}. 
These parameters are good milestones in summarizing the experimentally obtained $H_{\rm c2}(T)$ curves and in discussing the two-band superconductivity. 

In this work, we present experimental and calculated results of the $H_{\rm c2}(T)$ phase diagram of various MgB$_2$ thin films on some different substrates prepared by the molecular beam epitaxy (MBE), the multiple-targets sputtering, and co-evaporating deposition apparatus. 
We measured magnetoresistance in the whole region of $H_{\rm c2}(T)$ curves by using a pulsed magnet up to 37 T. 
In order to obtain the best fitting parameters of the Gurevich theory to experimental data, we newly took a differential plot, $(-dH_{\rm c2}/dT)$ $vs$. $T$, because this plot emphasizes the temperature dependence of the $H_{\rm c2}(T)$ curve. 
We find that the parameters obtained from the analysis are strongly correlated to the $T_{\rm c}$ of the films. 
Moreover, we present another interesting result on the transition width. 
In the process of magnetoresistance measurements at several temperatures, we found that the resistive transition width does not monotonically increase with decreasing temperature but has a minimum at a certain temperature below $T_{\rm c}$. 
We discuss the peculiar behavior in the temperature dependence of the transition width obtained from magnetoresistance measurements. 

\section{Experiment}
\begin{figure}
\includegraphics[trim=0mm 0mm 0mm 0mm, clip, height=5cm]{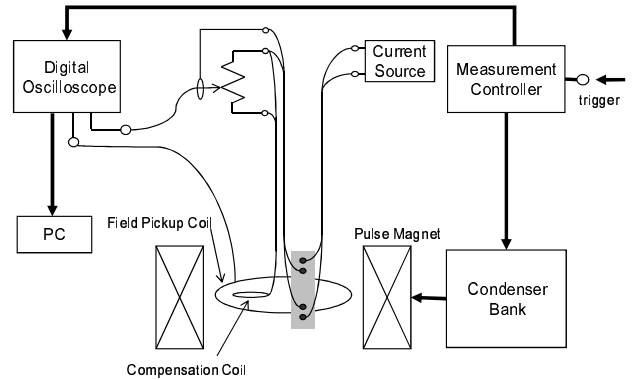}
\caption{\label{f01} Block diagram for the $H_{\rm c2}$ measurements under a pulsed magnetic field.}
\end{figure}

$H_{\rm c2}$ was determined from measuring the magnetoresistance by dc four-terminal method at several temperatures below $T_{\rm c}$ down to 1.5 K under a pulsed magnetic field up to 37 T using a home-made pulsed magnet system. 
Magnetoresistance was measured in the field direction of both $H \parallel ab$-plane and $H \parallel c$-axis to obtain superconducting anisotropy of the films. 
A block diagram for the $H_{\rm c2}$ measurements is shown in Fig.~\ref{f01}. 
In order to avoid excessive electric noise and heat-inflow to the sample, a battery is used as a current source in a proper electric circuit. 
Currents of 1-40 mA are applied to MgB$_2$ films, depending on the resistance of the samples, to obtain about 3 mV signal from the normal resistance. 
The magnetic field is applied by triggering a thyristor switch of LCR circuit from a delay pulse generator. 
The pulsed width of magnetic field is 10 ms. 
A field trace is obtained by integrating the voltage data recorded in a digital oscilloscope from a field pick-up coil. 
A stray signal from the lead wires is cancelled by using a compensation coil and a bridge circuit. 

Six as-grown MgB$_2$ thin films were used in the magnetoresistance measurements. 
Four MgB$_2$ films were prepared by Iwate group using an MBE apparatus with the co-evaporation conditions of low deposition rate in ultra-high vacuum without post annealing process \cite{harada2005, takahashi2006, harada2007IEEE}. 
Two of them were deposited on MgO (100) (\#1) and Si (111) (\#2) substrates at 200 $^\circ$C, the others were deposited on Ti buffer layers on ZnO (0001) substrates at 200 $^\circ$C. 
The difference of the latter two films is the thickness of the Ti buffer layers; one is 10 nm (\#3) and the other is 50 nm (\#4). 
The film thickness and the resistivity at 40 K are listed in Table~\ref{t1}. 
The structure and crystallinity were checked by x-ray diffraction. 
It was revealed that the $c$-axis of the MgB$_2$ orients a perpendicular direction to the film surface. 
Especially, the film \#4 is found to have an excellent alignment with the in-plane orientation \cite{harada2007IEEE}. 

\begin{figure}
\includegraphics[trim=0mm 0mm 0mm 0mm, clip, height=2.5cm]{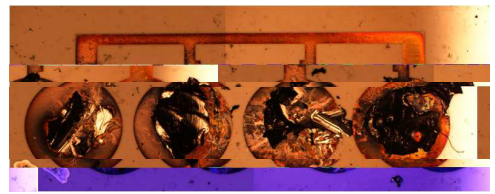}
\caption{\label{f02} Photograph of MgB$_2$ thin film \#6 used in the $H_{\rm c2}$ measurements. The film width is 0.2 mm and the length between terminals is 2 mm. }
\end{figure}

The other two MgB$_2$ films were prepared by NICT group. 
They were deposited on the $c$-plane of a sapphire substrate: one is fabricated by using a carrousel-type multiple-targets sputtering system without any buffer layers \cite{saito2002} denoted as \#5, and the other is by using a co-evaporation method (\#6)\cite{shimakage2004}.
They have a smooth surface as grown at low-substrate temperatures and their $c$-axis also well orients a perpendicular direction to the film surface. 
The film thickness and the resistivity at 40 K are also listed in Table~\ref{t1}. 
The co-evaporation film \#6 was fabricated to the shape of a four-terminal pattern as shown in Fig.~\ref{f02}. 

\begin{table}
\caption{\label{t1}Fabrication conditions of MgB$_2$ thin films. }
\begin{ruledtabular}
\begin{tabular}{ccccc}
No.&thickness&$\rho$(40K)&substrates&apparatus\\
&(nm)&($\mu\Omega$cm)&&\\
\hline
 \#1&100&9.2&MgO(100)&MBE\\
 \#2&300&35&Si(111)&MBE\\
 \#3&200&40&ZnO(0001)+10nm Ti&MBE\\
 \#4&200&4&ZnO(0001)+50nm Ti&MBE\\
 \#5&160&270&$c$-Al$_2$O$_3$&Sputter\\
 \#6&700&44&$c$-Al$_2$O$_3$&Co-evaporation\\
\end{tabular}
\end{ruledtabular}
\end{table}

\section{Results}
\begin{figure}
\includegraphics[trim=0mm 0mm 0mm 0mm, clip, height=9cm]{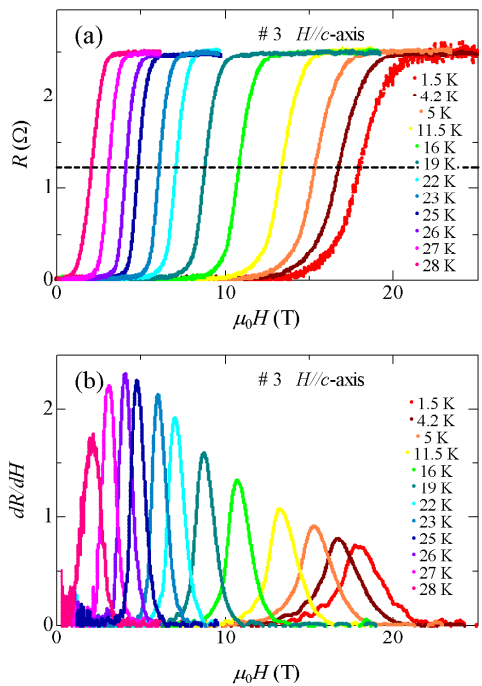}
\caption{\label{f03} (a) Magnetoresistance curves of MgB$_2$ film \#3 for $H \parallel c$-axis at several temperatures. (b) Differential plot of (a).}
\end{figure}

\begin{figure*}
\includegraphics[trim=0mm 0mm 0mm 0mm, clip, height=9cm]{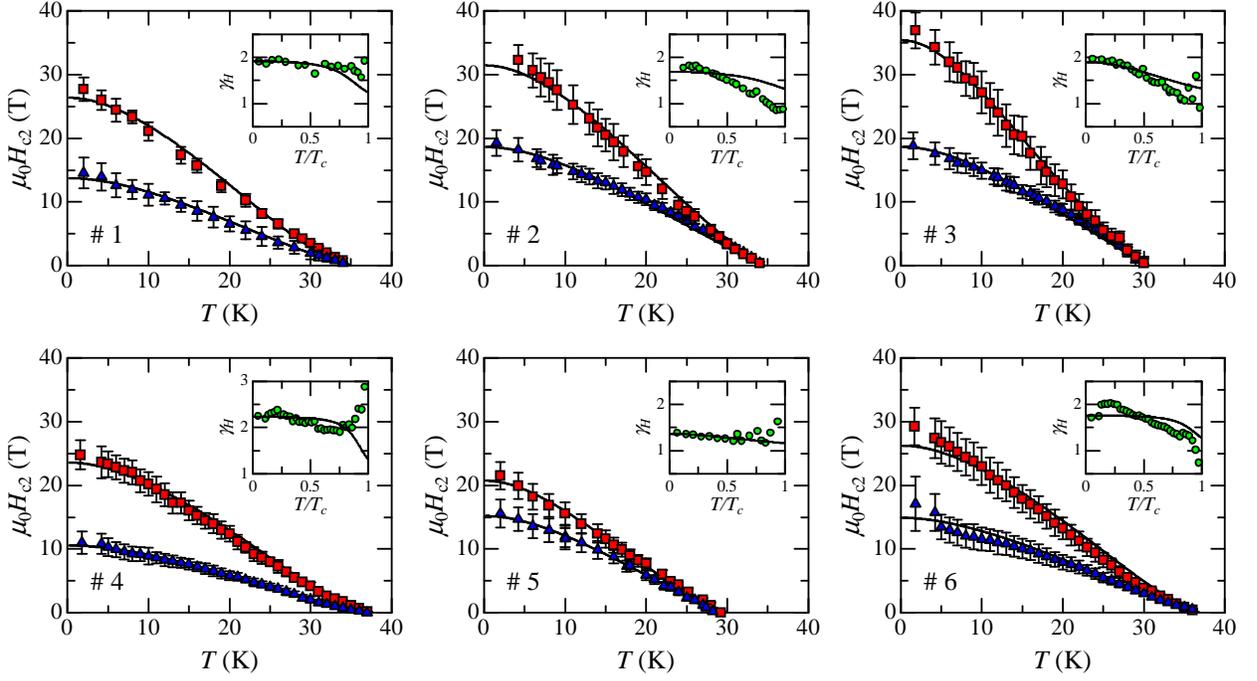}
\caption{\label{f04}$H_{\rm c2}(T)$ curves of six films for the field direction $H \parallel ab$-plane (square) and $H \parallel c$-axis (triangle). Insets show temperature dependence of anisotropy parameter. Solid lines are calculation curves obtained from fitting the Gurevich theory. }
\end{figure*}

The magnetoresistance curves of the MgB$_2$ film \#3 at several temperatures for $H \parallel c$-axis are shown in Fig.~\ref{f03}~(a) as a typical example of the experimental data. 
No hysteresis is observed. The $H_{\rm c2}$'s were defined as the midpoint of the resistive transition and the transition widths were defined as the fields between 10\% and 90\% of the normal resistance. 
The differential data $dR/dH$ are plotted as a function of the magnetic field in Fig.~\ref{f03}~(b), where the peak fields coincide with the $H_{\rm c2}$'s at each temperature. 
The peak width, which are almost proportional to the inverse of the peak heights, is also corresponding to the transition width. 
So, a maximum in the peak height observed at 26 K means that the transition becomes sharp at that temperature. 

From the magnetoresistance measurements, we obtained the $H_{\rm c2}(T)$ phase diagrams of six films, which are shown in Fig.~\ref{f04}. 
Error bars in the figure correspond to the fields between 10\% and 90\% of the normal resistance. 
Superconducting anisotropy parameter $\gamma_H = H_{\rm c2}^{(ab)}/H_{\rm c2}^{(c)}$ as a function of temperature was obtained as shown in the inset of the figure, where $H_{\rm c2}^{(ab)}$ and $H_{\rm c2}^{(c)}$ are the $H_{\rm c2}$ for $H \parallel ab$-plane and $H \parallel c$-axis, respectively. 
The $\mu_0 H_{\rm c2}^{(ab)}(0)$'s attain to 20$\sim$40 T, which are apparently larger than those for MgB$_2$ single crystal. 
This indicates that the superconducting coherence length is effectively reduced through the reduction of the electron mean free path due to some kind of disorder in fabricating the films. 
As for the anisotropy, the values of $\gamma_H$ are between one and three depending on the films, which are smaller than that of single crystal MgB$_2$. 
$\gamma_H$ decreases with increasing temperature, as shown in the inset of the figure. 
This is a basis for determining the ratio of the diffusivities for $\sigma$ and $\pi$ band as analyzed later. 
These experimentally obtained results of $T_{\rm c}$, $\mu_0H_{\rm c2}^{(ab)}(0)$, $\mu_0H_{\rm c2}^{(c)}(0)$ and $\gamma_H(0)$ are listed in Table~\ref{t2}. 

\section{Analysis}
\begin{figure}
\includegraphics[trim=0mm 0mm 0mm 0mm, clip, height=5.5cm]{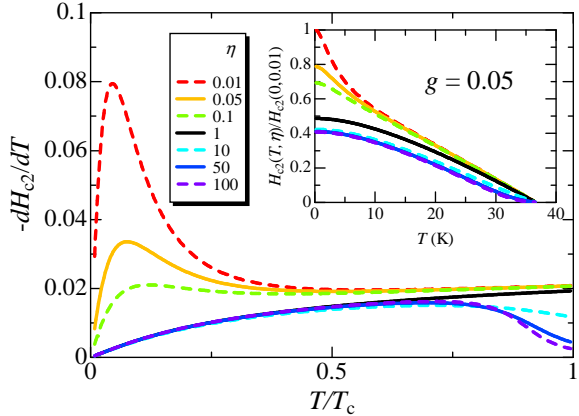}
\caption{\label{f05} Calculation curves of the Gurevich theory in $-dH_{\rm c2}/dT$ {\it vs.} $T/T_{\rm c}$ plot under several values of aparameter $\eta$. A parameter $g$ is fixed to be 0.05. Inset shows the calculated $H_{\rm c2}(T)$ curves with the same parameters. }
\end{figure}

\begin{figure*}
\includegraphics[trim=0mm 0mm 0mm 0mm, clip, height=9cm]{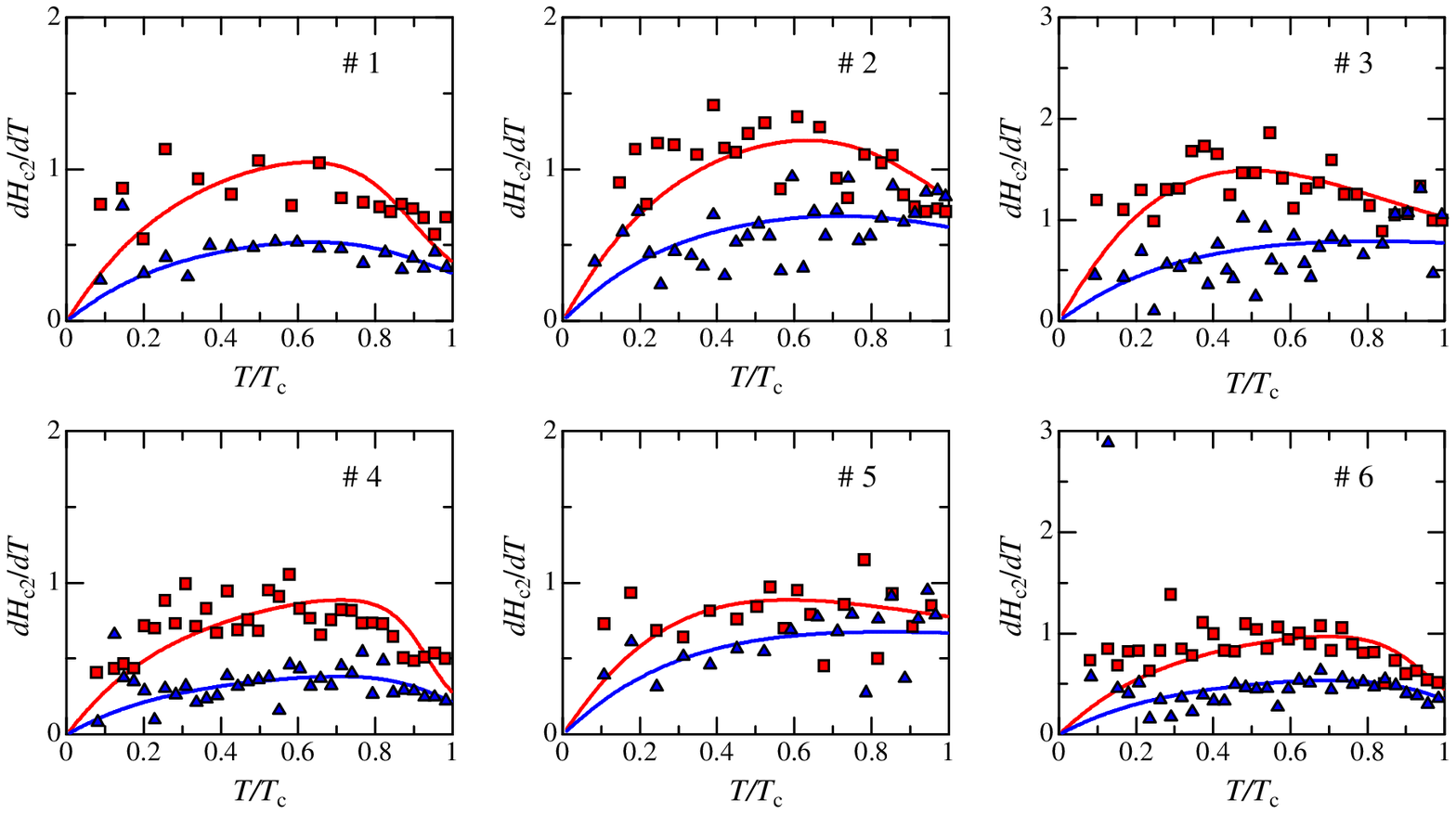}
\caption{\label{f06}$dH_{\rm c2}/dT$ $vs$. $T$ plots of six films for the field direction $H \parallel ab$-plane and $H \parallel c$-axis. Solid lines are fitting curves by using the Gurevich theory with the same parameters as shown in Fig.~\ref{f04}.}
\end{figure*}

Experimentally obtained data of the $H_{\rm c2}(T)$ and $\gamma_H$ were calculated by fitting the Gurevich theory \cite{gurevich2003,gurevich2007, braccini2005} based on two-gap Usadel equations, in which impurity scattering is introduced as the intraband electron diffusivities and the interband scattering. 
The calculations were performed using an application software (Mathematica, Wolfram Research Co.). 
For two-gap superconductivity, $T_{\rm c}$ is not affected by intraband scattering, but decreased with increasing the interband scattering parameter $g$. 
We assume $T_{\rm c0}$ = $T_{\rm c}(g = 0)$ = 40 K and then determine the value of $g$ from the Gurevich theory \cite{gurevich2007} as follows, 
\begin{equation}
U \left( \frac{g}{t_{\rm c}} \right) + \frac{(\lambda_0 + w {\rm ln}t_{\rm c}){\rm ln}t_{\rm c}}{p + w {\rm ln}t_{\rm c}} = 0,
\end{equation}
\begin{eqnarray}
U(x) = \psi (1/2+x) - \psi (1/2),\\
2p = \lambda_0 + [\lambda_{-}\gamma_{-}-2\lambda_{\pi\sigma}\gamma_{\sigma\pi}-2\lambda_{\sigma\pi}\gamma_{\pi\sigma}]/\gamma_{+},\\
\gamma_{\pm}=\gamma_{\sigma\pi}\pm\gamma_{\pi\sigma},\\
\lambda_{\pm}=\lambda_{\sigma\sigma}\pm\lambda_{\pi\pi},\\
\lambda_0=(\lambda_{-}^2+4\lambda_{\sigma\pi}\lambda_{\pi\sigma})^{1/2},\\
w=\lambda_{\sigma\sigma}\lambda_{\pi\pi}-\lambda_{\sigma\pi}\lambda_{\pi\sigma},\\
g=\hbar \gamma_{+}/2\pi k_{\rm B} T_{\rm c0},\\
t_{\rm c}=T_{\rm c}/T_{\rm c0},
\end{eqnarray}
where $\psi (x)$ is a digamma function. 
$\lambda_{\sigma\sigma}$ and $\lambda_{\pi\pi}$ are the intraband superconducting coupling constants of the $\sigma$ and $\pi$ bands, respectively. 
$\lambda_{\sigma\pi}$ and $\lambda_{\pi\sigma}$ are the interband superconducting coupling constants and $\gamma_{\sigma\pi}$ and $\gamma_{\pi\sigma}$ are the interband scattering rates. 
There is a constraint among these interband parameters so as to satisfy the symmetry relation to the partial densities of states $N_{\sigma}$ and $N_{\pi}$ in the $\sigma$ and $\pi$ band, respectively: 
\begin{eqnarray}
\frac{\lambda_{\sigma\pi}}{\lambda_{\pi\sigma}}=\frac{\gamma_{\sigma\pi}}{\gamma_{\pi\sigma}}=\frac{N_{\pi}}{N_{\sigma}},
\end{eqnarray}
where $N_{\pi}/N_{\sigma} \sim 1.3$ for MgB$_2$ \cite{gurevich2003}. 
We use the values of $\lambda_{\sigma\sigma}$, $\lambda_{\pi\pi}$, $\lambda_{\sigma\pi}$ and $\lambda_{\pi\sigma}$ fixed to be 0.81, 0.285, 0.119 and 0.09, respectively in the following calculations \cite {gurevich2007, golubov2002}. 
In this way, we determined the values of $g$ for six films using the experimentally obtained $T_{\rm c}$, as listed in Table~\ref{t2}. 

The equation for calculating $H_{\rm c2}$ is presented as follows, 
\begin{eqnarray}
(\lambda_0 + \lambda_i)[{\rm ln}t + U(x_{+})]+(\lambda_0 - \lambda_i)[{\rm ln}t + U( x_{-})]\nonumber\\
+ 2w[{\rm ln}t + U( x_{+})][{\rm ln}t + U( x_{-})] = 0,
\end{eqnarray}
\begin{eqnarray}
\lambda_i = [(\omega_{-} + \gamma_{-})\lambda_{-} -2\lambda_{\sigma\pi}\gamma_{\pi\sigma}-2\lambda_{\pi\sigma}\gamma_{\sigma\pi}]/\Omega_0,\\
x_{\pm} = \hbar (\omega_{+} + \gamma_{+} \pm \Omega_0)/4\pi k_{\rm B}T,\\
\Omega_0 = [(\omega_{-} + \gamma_{-})^2+4\gamma_{\sigma\pi}\gamma_{\pi\sigma}]^{1/2},\\
\omega_{\pm} = (D_{\sigma} \pm D_{\pi})\pi \mu_0 H/\phi_0,
\label{e01}
\end{eqnarray}
where $t = T/T_{\rm c0}$ and $\phi_0$ is the flux quantum. 
$D_{\sigma}$ and $D_{\pi}$ are the intraband electron diffusivities in the $\sigma$ and $\pi$ bands, respectively. 
Since the $\pi$ band has three-dimensional nature much more than the $\sigma$ band, we assume isotropic diffusivity in the $\pi$ band; $D_{\pi} = D_{\pi}^{(ab)} = D_{\pi}^{(c)}$. 
In the $\sigma$ band, on the other hand, anisotropic diffusivities along $c$-axis and $ab$-plane are introduced; $D_{\sigma} = D_{\sigma}^{(ab)}$ for $H \parallel c$-axis, while $D_{\sigma} = \sqrt{D_{\sigma}^{(ab)} D_{\sigma}^{(c)}}$ for $H \parallel ab$-plane. 
$D_{\sigma}^{(ab)}$ and $D_{\sigma}^{(c)}$ are the diffusivities along the $ab$-plane and the $c$-axis, respectively. 
The ratio $D_{\pi}/D_{\sigma} = \eta$ is defined as $\eta^{(ab)}= D_{\pi}/\sqrt{D_{\sigma}^{(ab)} D_{\sigma}^{(c)}}$ for $H \parallel ab$-plane and $\eta^{(c)}= D_{\pi}/D_{\sigma}^{(ab)}$ for $H \parallel c$-axis, respectively. 

In the calculation, the absolute value of $D_{\sigma}$ is roughly determined from the value of the $H_{\rm c2}(0)$ of each field direction. 
On the other hand, the shape of the $H_{\rm c2}(T)$ curve is mainly dependent on the parameter $\eta$. 
In the case of $\eta << 1$, the curve shows a steep increase at a low temperature near 0 K, while in the case of $\eta >> 1$, the initial slope of the $H_{\rm c2}(T)$ curve at $T_{\rm c}$ is suppressed as shown in the inset of Fig.~\ref{f05}. 
As $\eta = 1$, the $H_{\rm c2}(T)$ curve is identical to the curve for a single-band superconductor. 
This situation is much clear in the differential plot, $-dH_{\rm c2}/dT$ $vs.$ $T/T_{\rm c}$, as shown in Fig.~\ref{f05}. 
Therefore, in fitting the $H_{\rm c2}(T)$ curves, we also referred the $-dH_{\rm c2}/dT$ plot as shown in Fig.~\ref{f06}. 
It becomes quite valuable in these advanced analyses that we measured the magnetoresistance at a lot of temperatures for each film. 
In this way, we successfully obtain the calculation curves drawn by solid lines in Fig.~\ref{f04} and Fig.~\ref{f06}, which show fairly good agreement with the experimental data points. 
Obtained parameters are listed in Table~\ref{t2}. 

\begin{table*}
\caption{\label{t2}Experimentally obtained values of some superconducting characteristics and the best fitting parameters of Gurevich calculation to the experimental data.}
\begin{ruledtabular} 
\begin{tabular}{cccccccccccc}
 &\multicolumn{4}{c}{experimental}&&\multicolumn{6}{c}{fitting parameters}\\
 Film No.&$T_{\rm c}$&$\mu_0H_{\rm c2}^{(ab)}(0)$&$\mu_0H_{\rm c2}^{(c)}(0)$&$\gamma_H(0)$&&$g$&$D_{\pi}$&$D_{\sigma}^{(ab)}$&$D_{\sigma}^{(c)}$&$\eta^{(ab)}$&$\eta^{(c)}$\\
 &(K)&(T)&(T)&&&&(cm$^2$/s)&(cm$^2$/s)&(cm$^2$/s)&&\\ \hline
 \#1&35.1&29.1&15.3&1.91&&0.077&20.88&1.48&0.38&27.7&14.1\\
 \#2&34.5&35.4&20.0&1.78&&0.089&6.92&1.08&0.35&11.3&6.4\\
 \#3&30.4&38.6&19.5&1.98&&0.197&3.85&0.70&0.11&13.8&5.5\\
 \#4&37.1&25.3&11.4&2.23&&0.042&44.16&2.28&0.45&43.9&19.4\\
 \#5&28.8&23.0&16.3&1.41&&0.265&4.20&0.65&0.19&12.1&6.5\\
 \#6&36.9&30.3&18.6&1.62&&0.046&22.78&1.60&0.50&25.4&14.3\\
\end{tabular}
\end{ruledtabular}
\end{table*}

\section{Discussions}
\begin{figure}
\includegraphics[trim=0mm 0mm 0mm 0mm, clip, height=7cm]{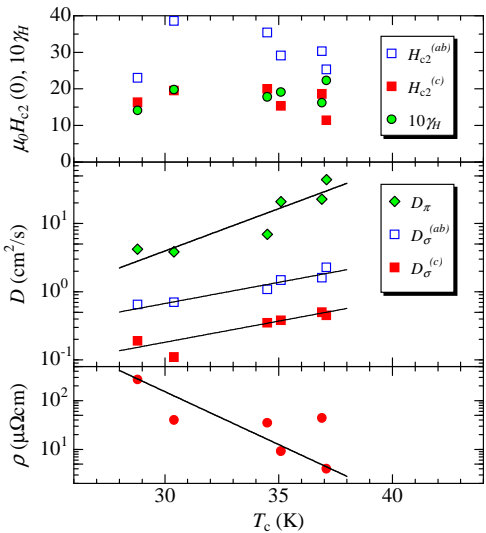}
\caption{\label{f07} Plots of several parameters obtained from this study against the $T_{\rm c}$.}
\end{figure}

The results of the analysis as listed in Table~\ref{t2} show the case of $\eta > 1$ for all MgB$_2$ films, which means the films have the cleaner $\pi$ band than the $\sigma$ band. 
This is confirmed also in the temperature dependence of $\gamma_H$ as shown in the insets of Fig.~\ref{f04}, where $\gamma_H$ decreases with increasing temperature, which is explained as the case of $\eta > 1$ in the Gurevich theory~\cite{gurevich2003}. 
There are many reports of the disorder effect on $H_{\rm c2}$ by doping a nonmagnetic impurity to MgB$_2$ bulks and films. 
For example, the $H_{\rm c2}$ attains up to 60 T in the C substitution system \cite{braccini2005}. 
On the contrary, our films were fabricated with no artificially doped impurity to get high grade films for device applications. 
Nevertheless, even such films show an enhancement of the $H_{\rm c2}$, which suggests that the structural disorder is more dominant than the substitution of nonmagnetic impurity atoms. 
Several factors to cause the structural disorder are deviation of the stoichiometry, mismatch of the lattice constants between MgB$_2$ and substrates (which causes a strain or distribution of the interatomic distance), randomness of in-plane orientation, degradation by the oxygen and water attack, and so on. 
It is quite natural that the structural disorder is equally introduced in both $\sigma$ and $\pi$ bands as well as in the interband scattering. 

In order to make clear the disorder effect, we plot several parameters obtained from this study against the $T_{\rm c}$ as shown in Fig.~\ref{f07}. 
The experimental parameters, $\mu_0H_{\rm c2}^{(ab)}(0)$, $\mu_0H_{\rm c2}^{(c)}(0)$ and $\gamma_H$ exhibit no $T_{\rm c}$ dependence as shown in the upper graph of Fig.~\ref{f07}. 
On the other hand, obtained parameters from the analysis, $D_{\pi}$, $D_{\sigma}^{(ab)}$ and $D_{\sigma}^{(c)}$ decrease clearly with decreasing $T_{\rm c}$. 
$D_{\sigma}^{(ab)}$ and $D_{\sigma}^{(c)}$ show an exponential dependence on the $T_{\rm c}$, as drawn by solid lines in the middle graph of Fig.~\ref{f07}. 
$D_{\pi}$ decreases much faster than $D_{\sigma}^{(ab)}$ and $D_{\sigma}^{(c)}$ with decreasing $T_{\rm c}$. 
Also, $\eta^{(ab)}$ and $\eta^{(c)}$ show similar $T_{\rm c}$ dependence to the diffusivity, though they are not plotted in Fig.~\ref{f07}. 
As for the resistivity, $\rho$ increases roughly with decreasing $T_{\rm c}$ as shown in the bottom graph of the figure. 
As a conclusion, $T_{\rm c}$ of the films is a good parameter to characterize the quality of the superconducting films even in two-band superconductors, as well as the residual resistivity, although there are many kinds of disorders on fabricating the MgB$_2$ films, such as non-stoichiometry, mismatch of the lattice constants between substrates and MgB$_2$, atmosphere or degree of vacuum, degradation by oxygen or water, and so on. 

\begin{table}
\caption{\label{t3}Several parameters fitting the Gurevich calculation to the experimental data of MgB$_2$ film \#3 for different definitions of the $H_{\rm c2}$. }
\begin{ruledtabular}
\begin{tabular}{cccccc}
\#3&$T_{\rm c}$&$g$&$D_{\pi}$&$D_{\sigma}^{(ab)}$&$D_{\sigma}^{(c)}$\\
 &(K)&&(cm$^2$/s)&(cm$^2$/s)&(cm$^2$/s)\\ \hline
midpoint&30.4&0.197&3.85&0.70&0.11\\
90\%$R_n$&31.3&0.168&4.05&0.72&0.14\\
99\%$R_n$&32.4&0.137&6.19&0.68&0.20\\
\end{tabular}
\end{ruledtabular}
\end{table}

Next, we discuss the definition of the $H_{\rm c2}$. 
The $H_{\rm c2}$ defined as an onset of the magnetoresistive transition corresponds to a thermodynamical $H_{\rm c2}$ as usually argued in the theoretical treatments. 
However, the onset fields determined experimentally are less relevant due to defocus of the transition than the midpoint fields which are also coincident with the peak fields of the $dR/dH$ plots as mentioned before. 
Therefore, the midpoint definition is much relevant from an experimental point of view in the systematic measurements and characterizations of various MgB$_2$ thin films fabricated by different methods \cite{kunchur2003}. 
Nevertheless, it is worth to check how obtained parameters from the calculation fitting to the experimental data are varied by the difference of the $H_{\rm c2}$ definitions. 
The magnetoresistance curves of the film \#3 are rather sharp as shown in Fig.~\ref{f03}, so we obtained each $H_{\rm c2}(T)$ determined from midpoint, 90\% and 99\% of the normal resistance ($R_n$). 
Although the $T_{\rm c}$'s and the $H_{\rm c2}(0)$'s are different according to the definition, the temperature dependence of the $H_{\rm c2}$ among three definitions is very similar to one another. 
Then, we also analyzed each data of the $H_{\rm c2}(T)$ by fitting the Gurevich theory. 
Obtained parameters are listed in Table~\ref{t3}. 
The parameters $D_{\pi}$, $D_{\sigma}^{(ab)}$ and $D_{\sigma}^{(c)}$ for 90\%$R_n$ are almost the same as those for the midpoint definition. 
On the other hand, $D_{\pi}$ and $D_{\sigma}^{(c)}$ for 99\%$R_n$ are about 1.5 times larger than those for the others. 
However, when we plot these data to Fig.~\ref{f07}, we find that they are plotted near the solid line of the figure. they are almost plotted on the line drawn in Fig.~\ref{f07}. 
Accordingly, the conclusion of the cleaner $\pi$ band is not changed by the difference of the $H_{\rm c2}$ definition. 

Finally, we discuss briefly the transition width in the magnetoresistance measurements. 
There are no discussions on the transition width in the two-band superconductivity as far as we know. 
As shown in Fig.~\ref{f03}, the magnetoresistive transition becomes sharp at 26 K and broad below that temperature. 
If we interpret this behavior only in terms of the $H_{\rm c2}(T)$ phase diagram, it may be related to them that the $H_{\rm c2}(T)$ curve shows an upward curvature. 
When the superconductivity in the $\pi$ band is destroyed by applied field, the normal core appears partly to form a vortex pinning center. 
The pinning mechanism of MgB$_2$ films has been speculatively discussed from the mixed-state transport measurements by Arcos and Kunchur \cite{arcos2005}. 
Although the transition width is often related to the local $T_{\rm c}$ variations due to some inhomogeneties, it may be attributed to two-band superconducting characteristics. 
Here, we point out that the transition width observed in some MgB$_2$ thin films does not show monotonic temperature dependence \cite{noguchi2008}. 

\section{Conclusion}
We presented experimental results on the $H_{\rm c2}(T)$ for $H \parallel ab$-plane and $H \parallel c$-axis of various as-grown MgB$_2$ thin films prepared by the molecular beam epitaxy, multiple-targets sputtering, and co-evaporation deposition apparatus. 
The results were well described by the Gurevich theory of dirty two-gap superconductivity. 
We extracted empirical parameters on the diffusivity of superconducting electrons for the $\sigma$ and $\pi$ bands from the $H_{\rm c2}(T)$ calculation. 
All films were categorized to $D_{\pi}/D_{\sigma}>1$, namely the cleaner $\pi$ band case. 
We found that the parameters obtained from the analysis are strongly correlated to the $T_{\rm c}$ of the films. 
Accordingly, $T_{\rm c}$ is a good parameter for characterizing the quality of the films in a view of superconducting properties, as well as the residual resistivity. 

\begin{acknowledgments}
We would like to thank A. Gurevich for critical reading of this manuscript. 
This work was partly supported by a Grant-in-Aid for Scientific Research from the Ministry of Education, Culture, Sports, Science and Technology of Japan (Grant No. 19206104). 
\end{acknowledgments}


\begin{thebibliography}{25}
\expandafter\ifx\csname natexlab\endcsname\relax\def\natexlab#1{#1}\fi
\expandafter\ifx\csname bibnamefont\endcsname\relax
  \def\bibnamefont#1{#1}\fi
\expandafter\ifx\csname bibfnamefont\endcsname\relax
  \def\bibfnamefont#1{#1}\fi
\expandafter\ifx\csname citenamefont\endcsname\relax
  \def\citenamefont#1{#1}\fi
\expandafter\ifx\csname url\endcsname\relax
  \def\url#1{\texttt{#1}}\fi
\expandafter\ifx\csname urlprefix\endcsname\relax\def\urlprefix{URL }\fi
\providecommand{\bibinfo}[2]{#2}
\providecommand{\eprint}[2][]{\url{#2}}

\bibitem[{\citenamefont{Nagamatsu et~al.}(2001)\citenamefont{Nagamatsu,
  Nakagawa, Muranaka, Zenitani, and Akimitsu}}]{nagamatsu2001}
\bibinfo{author}{\bibfnamefont{J.}~\bibnamefont{Nagamatsu}},
  \bibinfo{author}{\bibfnamefont{N.}~\bibnamefont{Nakagawa}},
  \bibinfo{author}{\bibfnamefont{T.}~\bibnamefont{Muranaka}},
  \bibinfo{author}{\bibfnamefont{Y.}~\bibnamefont{Zenitani}}, \bibnamefont{and}
  \bibinfo{author}{\bibfnamefont{J.}~\bibnamefont{Akimitsu}},
  \bibinfo{journal}{Nature} \textbf{\bibinfo{volume}{410}}, \bibinfo{pages}{63}
  (\bibinfo{year}{2001}).

\bibitem[{\citenamefont{Choi et~al.}(2002{\natexlab{a}})\citenamefont{Choi,
  Roundy, Sun, Cohen, and Louie}}]{choi2002nature}
\bibinfo{author}{\bibfnamefont{H.~J.} \bibnamefont{Choi}},
  \bibinfo{author}{\bibfnamefont{D.}~\bibnamefont{Roundy}},
  \bibinfo{author}{\bibfnamefont{H.}~\bibnamefont{Sun}},
  \bibinfo{author}{\bibfnamefont{M.~L.} \bibnamefont{Cohen}}, \bibnamefont{and}
  \bibinfo{author}{\bibfnamefont{S.~G.} \bibnamefont{Louie}},
  \bibinfo{journal}{Nature} \textbf{\bibinfo{volume}{418}},
  \bibinfo{pages}{758} (\bibinfo{year}{2002}{\natexlab{a}}).

\bibitem[{\citenamefont{Choi et~al.}(2002{\natexlab{b}})\citenamefont{Choi,
  Roundy, Sun, Cohen, and Louie}}]{choi2002PRB}
\bibinfo{author}{\bibfnamefont{H.~J.} \bibnamefont{Choi}},
  \bibinfo{author}{\bibfnamefont{D.}~\bibnamefont{Roundy}},
  \bibinfo{author}{\bibfnamefont{H.}~\bibnamefont{Sun}},
  \bibinfo{author}{\bibfnamefont{M.~L.} \bibnamefont{Cohen}}, \bibnamefont{and}
  \bibinfo{author}{\bibfnamefont{S.~G.} \bibnamefont{Louie}},
  \bibinfo{journal}{Phys. Rev. B} \textbf{\bibinfo{volume}{66}},
  \bibinfo{pages}{020513(R)} (\bibinfo{year}{2002}{\natexlab{b}}).

\bibitem[{\citenamefont{Golubov et~al.}(2002)\citenamefont{Golubov, Kortus,
  Dolgov, Jepsen, Kong, Andersen, Gibson, Ahn, and Kremer}}]{golubov2002}
\bibinfo{author}{\bibfnamefont{A.~A.} \bibnamefont{Golubov}},
  \bibinfo{author}{\bibfnamefont{J.}~\bibnamefont{Kortus}},
  \bibinfo{author}{\bibfnamefont{O.~V.} \bibnamefont{Dolgov}},
  \bibinfo{author}{\bibfnamefont{O.}~\bibnamefont{Jepsen}},
  \bibinfo{author}{\bibfnamefont{Y.}~\bibnamefont{Kong}},
  \bibinfo{author}{\bibfnamefont{O.~K.} \bibnamefont{Andersen}},
  \bibinfo{author}{\bibfnamefont{B.~J.} \bibnamefont{Gibson}},
  \bibinfo{author}{\bibfnamefont{K.}~\bibnamefont{Ahn}}, \bibnamefont{and}
  \bibinfo{author}{\bibfnamefont{R.~K.} \bibnamefont{Kremer}},
  \bibinfo{journal}{J. Phys. Cond. Mat.} \textbf{\bibinfo{volume}{14}},
  \bibinfo{pages}{1353} (\bibinfo{year}{2002}).

\bibitem[{\citenamefont{Golubov and Koshelev}(2003)}]{golubov2003}
\bibinfo{author}{\bibfnamefont{A.~A.} \bibnamefont{Golubov}} \bibnamefont{and}
  \bibinfo{author}{\bibfnamefont{A.~E.} \bibnamefont{Koshelev}},
  \bibinfo{journal}{Phys. Rev. B} \textbf{\bibinfo{volume}{68}},
  \bibinfo{pages}{104503} (\bibinfo{year}{2003}).

\bibitem[{\citenamefont{Gurevich}(2003)}]{gurevich2003}
\bibinfo{author}{\bibfnamefont{A.}~\bibnamefont{Gurevich}},
  \bibinfo{journal}{Phys. Rev. B} \textbf{\bibinfo{volume}{67}},
  \bibinfo{pages}{184515} (\bibinfo{year}{2003}).

\bibitem[{\citenamefont{Gurevich et~al.}(2004)\citenamefont{Gurevich, Patnaik,
  Braccini, Kim, Mielke, Song, Cooley, Bu, Kim, Choi et~al.}}]{gurevich2004}
\bibinfo{author}{\bibfnamefont{A.}~\bibnamefont{Gurevich}},
  \bibinfo{author}{\bibfnamefont{S.}~\bibnamefont{Patnaik}},
  \bibinfo{author}{\bibfnamefont{V.}~\bibnamefont{Braccini}},
  \bibinfo{author}{\bibfnamefont{K.~H.} \bibnamefont{Kim}},
  \bibinfo{author}{\bibfnamefont{C.}~\bibnamefont{Mielke}},
  \bibinfo{author}{\bibfnamefont{X.}~\bibnamefont{Song}},
  \bibinfo{author}{\bibfnamefont{L.~D.} \bibnamefont{Cooley}},
  \bibinfo{author}{\bibfnamefont{S.~D.} \bibnamefont{Bu}},
  \bibinfo{author}{\bibfnamefont{D.~M.} \bibnamefont{Kim}},
  \bibinfo{author}{\bibfnamefont{J.~H.} \bibnamefont{Choi}},
  \bibnamefont{et~al.}, \bibinfo{journal}{Supercond. Sci. Technol.}
  \textbf{\bibinfo{volume}{17}}, \bibinfo{pages}{278} (\bibinfo{year}{2004}).

\bibitem[{\citenamefont{Braccini et~al.}(2005)\citenamefont{Braccini, Gurevich,
  Giencke, Jewell, Eom, Larbalestier, Pogrebnyakov, Cui, Liu, Hu
  et~al.}}]{braccini2005}
\bibinfo{author}{\bibfnamefont{V.}~\bibnamefont{Braccini}},
  \bibinfo{author}{\bibfnamefont{A.}~\bibnamefont{Gurevich}},
  \bibinfo{author}{\bibfnamefont{J.~E.} \bibnamefont{Giencke}},
  \bibinfo{author}{\bibfnamefont{M.~C.} \bibnamefont{Jewell}},
  \bibinfo{author}{\bibfnamefont{C.~B.} \bibnamefont{Eom}},
  \bibinfo{author}{\bibfnamefont{D.~C.} \bibnamefont{Larbalestier}},
  \bibinfo{author}{\bibfnamefont{A.}~\bibnamefont{Pogrebnyakov}},
  \bibinfo{author}{\bibfnamefont{Y.}~\bibnamefont{Cui}},
  \bibinfo{author}{\bibfnamefont{B.~T.} \bibnamefont{Liu}},
  \bibinfo{author}{\bibfnamefont{Y.~F.} \bibnamefont{Hu}},
  \bibnamefont{et~al.}, \bibinfo{journal}{Phys. Rev. B}
  \textbf{\bibinfo{volume}{71}}, \bibinfo{eid}{012504} (\bibinfo{year}{2005}).

\bibitem[{\citenamefont{Gurevich}(2007)}]{gurevich2007}
\bibinfo{author}{\bibfnamefont{A.}~\bibnamefont{Gurevich}},
  \bibinfo{journal}{Physica C} \textbf{\bibinfo{volume}{456}},
  \bibinfo{pages}{160} (\bibinfo{year}{2007}).

\bibitem[{\citenamefont{Kortus et~al.}(2001)\citenamefont{Kortus, Mazin,
  Belashchenko, Antropov, and Boyer}}]{kortus2001}
\bibinfo{author}{\bibfnamefont{J.}~\bibnamefont{Kortus}},
  \bibinfo{author}{\bibfnamefont{I.~I.} \bibnamefont{Mazin}},
  \bibinfo{author}{\bibfnamefont{K.~D.} \bibnamefont{Belashchenko}},
  \bibinfo{author}{\bibfnamefont{V.~P.} \bibnamefont{Antropov}},
  \bibnamefont{and} \bibinfo{author}{\bibfnamefont{L.~L.} \bibnamefont{Boyer}},
  \bibinfo{journal}{Phys. Rev. Lett.} \textbf{\bibinfo{volume}{86}},
  \bibinfo{pages}{4656} (\bibinfo{year}{2001}).

\bibitem[{\citenamefont{Liu et~al.}(2001)\citenamefont{Liu, Mazin, and
  Kortus}}]{liu2001}
\bibinfo{author}{\bibfnamefont{A.~Y.} \bibnamefont{Liu}},
  \bibinfo{author}{\bibfnamefont{I.~I.} \bibnamefont{Mazin}}, \bibnamefont{and}
  \bibinfo{author}{\bibfnamefont{J.}~\bibnamefont{Kortus}},
  \bibinfo{journal}{Phys.\ Rev.\ Lett.} \textbf{\bibinfo{volume}{87}},
  \bibinfo{pages}{087005} (\bibinfo{year}{2001}).

\bibitem[{\citenamefont{Lyard et~al.}(2002)\citenamefont{Lyard, Samuely, Szabo,
  Klein, Marcenat, Paulius, Kim, Jung, and Lee}}]{lyard2002}
\bibinfo{author}{\bibfnamefont{L.}~\bibnamefont{Lyard}},
  \bibinfo{author}{\bibfnamefont{P.}~\bibnamefont{Samuely}},
  \bibinfo{author}{\bibfnamefont{P.}~\bibnamefont{Szabo}},
  \bibinfo{author}{\bibfnamefont{T.}~\bibnamefont{Klein}},
  \bibinfo{author}{\bibfnamefont{C.}~\bibnamefont{Marcenat}},
  \bibinfo{author}{\bibfnamefont{L.}~\bibnamefont{Paulius}},
  \bibinfo{author}{\bibfnamefont{K.~H.~P.} \bibnamefont{Kim}},
  \bibinfo{author}{\bibfnamefont{C.~U.} \bibnamefont{Jung}}, \bibnamefont{and}
  \bibinfo{author}{\bibfnamefont{H.~S.} \bibnamefont{Lee}},
  \bibinfo{journal}{Phys. Rev. B} \textbf{\bibinfo{volume}{66}},
  \bibinfo{pages}{180502(R)} (\bibinfo{year}{2002}).

\bibitem[{\citenamefont{Jung et~al.}(2001)\citenamefont{Jung, Jaime, Lacerda,
  Boebinger, Kang, Kim, Choi, and Lee}}]{jung2001}
\bibinfo{author}{\bibfnamefont{M.~H.} \bibnamefont{Jung}},
  \bibinfo{author}{\bibfnamefont{M.}~\bibnamefont{Jaime}},
  \bibinfo{author}{\bibfnamefont{A.~H.} \bibnamefont{Lacerda}},
  \bibinfo{author}{\bibfnamefont{G.~S.} \bibnamefont{Boebinger}},
  \bibinfo{author}{\bibfnamefont{W.~N.} \bibnamefont{Kang}},
  \bibinfo{author}{\bibfnamefont{H.~J.} \bibnamefont{Kim}},
  \bibinfo{author}{\bibfnamefont{E.~M.} \bibnamefont{Choi}}, \bibnamefont{and}
  \bibinfo{author}{\bibfnamefont{S.~I.} \bibnamefont{Lee}},
  \bibinfo{journal}{Chem. Phys. Lett.} \textbf{\bibinfo{volume}{343}},
  \bibinfo{pages}{447} (\bibinfo{year}{2001}).

\bibitem[{\citenamefont{Ferrando et~al.}(2003)\citenamefont{Ferrando,
  Manfrinetti, Marre, Putti, Sheikin, Tarantini, and
  Ferdeghini}}]{ferrando2003}
\bibinfo{author}{\bibfnamefont{V.}~\bibnamefont{Ferrando}},
  \bibinfo{author}{\bibfnamefont{P.}~\bibnamefont{Manfrinetti}},
  \bibinfo{author}{\bibfnamefont{D.}~\bibnamefont{Marre}},
  \bibinfo{author}{\bibfnamefont{M.}~\bibnamefont{Putti}},
  \bibinfo{author}{\bibfnamefont{I.}~\bibnamefont{Sheikin}},
  \bibinfo{author}{\bibfnamefont{C.}~\bibnamefont{Tarantini}},
  \bibnamefont{and}
  \bibinfo{author}{\bibfnamefont{C.}~\bibnamefont{Ferdeghini}},
  \bibinfo{journal}{Phys. Rev. B} \textbf{\bibinfo{volume}{68}},
  \bibinfo{pages}{094517} (\bibinfo{year}{2003}).

\bibitem[{\citenamefont{Noguchi et~al.}(2005)\citenamefont{Noguchi, Miki,
  Shimakage, Wang, Satoh, Yotsuya, and Ishida}}]{noguchi2005}
\bibinfo{author}{\bibfnamefont{S.}~\bibnamefont{Noguchi}},
  \bibinfo{author}{\bibfnamefont{S.}~\bibnamefont{Miki}},
  \bibinfo{author}{\bibfnamefont{H.}~\bibnamefont{Shimakage}},
  \bibinfo{author}{\bibfnamefont{Z.}~\bibnamefont{Wang}},
  \bibinfo{author}{\bibfnamefont{K.}~\bibnamefont{Satoh}},
  \bibinfo{author}{\bibfnamefont{T.}~\bibnamefont{Yotsuya}}, \bibnamefont{and}
  \bibinfo{author}{\bibfnamefont{T.}~\bibnamefont{Ishida}},
  \bibinfo{journal}{Physica C} \textbf{\bibinfo{volume}{426-431}},
  \bibinfo{pages}{1449} (\bibinfo{year}{2005}).

\bibitem[{\citenamefont{Lawrence and Doniach}(1971)}]{lawrence1971}
\bibinfo{author}{\bibfnamefont{W.~E.} \bibnamefont{Lawrence}} \bibnamefont{and}
  \bibinfo{author}{\bibfnamefont{S.}~\bibnamefont{Doniach}},
  \bibinfo{journal}{Proc. of the 12th Int. Conf. on Low Temp. Phys., Kyoto,
  edited by E. Kanda (Keigaku, Tokyo, 1971)} p. \bibinfo{pages}{361}
  (\bibinfo{year}{1971}).

\bibitem[{\citenamefont{Helfand and Werthamer}(1964)}]{helfand1964}
\bibinfo{author}{\bibfnamefont{E.}~\bibnamefont{Helfand}} \bibnamefont{and}
  \bibinfo{author}{\bibfnamefont{N.~R.} \bibnamefont{Werthamer}},
  \bibinfo{journal}{Phys. Rev. Lett.} \textbf{\bibinfo{volume}{13}},
  \bibinfo{pages}{686} (\bibinfo{year}{1964}).

\bibitem[{\citenamefont{Harada et~al.}(2005)\citenamefont{Harada, Takahashi,
  Iriuda, Kuroha, Nakanishi, and Yoshizawa}}]{harada2005}
\bibinfo{author}{\bibfnamefont{Y.}~\bibnamefont{Harada}},
  \bibinfo{author}{\bibfnamefont{T.}~\bibnamefont{Takahashi}},
  \bibinfo{author}{\bibfnamefont{H.}~\bibnamefont{Iriuda}},
  \bibinfo{author}{\bibfnamefont{M.}~\bibnamefont{Kuroha}},
  \bibinfo{author}{\bibfnamefont{Y.}~\bibnamefont{Nakanishi}},
  \bibnamefont{and}
  \bibinfo{author}{\bibfnamefont{M.}~\bibnamefont{Yoshizawa}},
  \bibinfo{journal}{Physica C} \textbf{\bibinfo{volume}{426-431}},
  \bibinfo{pages}{1453} (\bibinfo{year}{2005}).

\bibitem[{\citenamefont{Takahashi et~al.}(2006)\citenamefont{Takahashi, Harada,
  Iriuda, Kuroha, Oba, Seki, Nakanishi, Echigoya, and
  Yoshizawa}}]{takahashi2006}
\bibinfo{author}{\bibfnamefont{T.}~\bibnamefont{Takahashi}},
  \bibinfo{author}{\bibfnamefont{Y.}~\bibnamefont{Harada}},
  \bibinfo{author}{\bibfnamefont{H.}~\bibnamefont{Iriuda}},
  \bibinfo{author}{\bibfnamefont{M.}~\bibnamefont{Kuroha}},
  \bibinfo{author}{\bibfnamefont{T.}~\bibnamefont{Oba}},
  \bibinfo{author}{\bibfnamefont{M.}~\bibnamefont{Seki}},
  \bibinfo{author}{\bibfnamefont{Y.}~\bibnamefont{Nakanishi}},
  \bibinfo{author}{\bibfnamefont{J.}~\bibnamefont{Echigoya}}, \bibnamefont{and}
  \bibinfo{author}{\bibfnamefont{M.}~\bibnamefont{Yoshizawa}},
  \bibinfo{journal}{Physica C} \textbf{\bibinfo{volume}{445-448}},
  \bibinfo{pages}{887} (\bibinfo{year}{2006}).

\bibitem[{\citenamefont{Harada et~al.}(2007)\citenamefont{Harada, Yamaguchi,
  Takahashi, Iriuda, Oba, and Yoshizawa}}]{harada2007IEEE}
\bibinfo{author}{\bibfnamefont{Y.}~\bibnamefont{Harada}},
  \bibinfo{author}{\bibfnamefont{H.}~\bibnamefont{Yamaguchi}},
  \bibinfo{author}{\bibfnamefont{T.}~\bibnamefont{Takahashi}},
  \bibinfo{author}{\bibfnamefont{H.}~\bibnamefont{Iriuda}},
  \bibinfo{author}{\bibfnamefont{T.}~\bibnamefont{Oba}}, \bibnamefont{and}
  \bibinfo{author}{\bibfnamefont{M.}~\bibnamefont{Yoshizawa}},
  \bibinfo{journal}{Applied Superconductivity, IEEE Transactions on}
  \textbf{\bibinfo{volume}{17}}, \bibinfo{pages}{2883} (\bibinfo{year}{2007}).

\bibitem[{\citenamefont{Saito et~al.}(2002)\citenamefont{Saito, Kawakami,
  Shimakage, and Wang}}]{saito2002}
\bibinfo{author}{\bibfnamefont{A.}~\bibnamefont{Saito}},
  \bibinfo{author}{\bibfnamefont{A.}~\bibnamefont{Kawakami}},
  \bibinfo{author}{\bibfnamefont{H.}~\bibnamefont{Shimakage}},
  \bibnamefont{and} \bibinfo{author}{\bibfnamefont{Z.}~\bibnamefont{Wang}},
  \bibinfo{journal}{Jpn. J. Appl. Phys.} \textbf{\bibinfo{volume}{41}},
  \bibinfo{pages}{L127} (\bibinfo{year}{2002}).

\bibitem[{\citenamefont{Shimakage et~al.}(2004)\citenamefont{Shimakage, Saito,
  Kawakami, and Wang}}]{shimakage2004}
\bibinfo{author}{\bibfnamefont{H.}~\bibnamefont{Shimakage}},
  \bibinfo{author}{\bibfnamefont{A.}~\bibnamefont{Saito}},
  \bibinfo{author}{\bibfnamefont{A.}~\bibnamefont{Kawakami}}, \bibnamefont{and}
  \bibinfo{author}{\bibfnamefont{Z.}~\bibnamefont{Wang}},
  \bibinfo{journal}{Physica C} \textbf{\bibinfo{volume}{408-410}},
  \bibinfo{pages}{891} (\bibinfo{year}{2004}).

\bibitem[{\citenamefont{Kunchur et~al.}(2003)\citenamefont{Kunchur, Wu, Arcos,
  Ivlev, Choi, Kim, Kang, and Lee}}]{kunchur2003}
\bibinfo{author}{\bibfnamefont{M.~N.} \bibnamefont{Kunchur}},
  \bibinfo{author}{\bibfnamefont{C.}~\bibnamefont{Wu}},
  \bibinfo{author}{\bibfnamefont{D.~H.} \bibnamefont{Arcos}},
  \bibinfo{author}{\bibfnamefont{B.~I.} \bibnamefont{Ivlev}},
  \bibinfo{author}{\bibfnamefont{E.-M.} \bibnamefont{Choi}},
  \bibinfo{author}{\bibfnamefont{K.~H.~P.} \bibnamefont{Kim}},
  \bibinfo{author}{\bibfnamefont{W.~N.} \bibnamefont{Kang}}, \bibnamefont{and}
  \bibinfo{author}{\bibfnamefont{S.-I.} \bibnamefont{Lee}},
  \bibinfo{journal}{Phys. Rev. B} \textbf{\bibinfo{volume}{68}},
  \bibinfo{pages}{100503(R)} (\bibinfo{year}{2003}).

\bibitem[{\citenamefont{Arcos and Kunchur}(2005)}]{arcos2005}
\bibinfo{author}{\bibfnamefont{D.~H.} \bibnamefont{Arcos}} \bibnamefont{and}
  \bibinfo{author}{\bibfnamefont{M.~N.} \bibnamefont{Kunchur}},
  \bibinfo{journal}{Phys. Rev. B} \textbf{\bibinfo{volume}{71}},
  \bibinfo{pages}{184516} (\bibinfo{year}{2005}).

\bibitem[{\citenamefont{Noguchi et~al.}(to be published in
  2008)\citenamefont{Noguchi, Kuribayashi, Harada, Yoshizawa, Miki, Shimakage,
  Wang, Satoh, Yotsuya, and Ishida}}]{noguchi2008}
\bibinfo{author}{\bibfnamefont{S.}~\bibnamefont{Noguchi}},
  \bibinfo{author}{\bibfnamefont{A.}~\bibnamefont{Kuribayashi}},
  \bibinfo{author}{\bibfnamefont{Y.}~\bibnamefont{Harada}},
  \bibinfo{author}{\bibfnamefont{M.}~\bibnamefont{Yoshizawa}},
  \bibinfo{author}{\bibfnamefont{S.}~\bibnamefont{Miki}},
  \bibinfo{author}{\bibfnamefont{H.}~\bibnamefont{Shimakage}},
  \bibinfo{author}{\bibfnamefont{Z.}~\bibnamefont{Wang}},
  \bibinfo{author}{\bibfnamefont{K.}~\bibnamefont{Satoh}},
  \bibinfo{author}{\bibfnamefont{T.}~\bibnamefont{Yotsuya}}, \bibnamefont{and}
  \bibinfo{author}{\bibfnamefont{T.}~\bibnamefont{Ishida}},
  \bibinfo{journal}{J. Phys. Chem. Solids}  (\bibinfo{year}{to be published in
  2008}).

\end{thebibliography}
\end{document}